\documentclass[prb,twocolumn,a4paper,showpacs,aps,superscriptaddress]{revtex4}
\usepackage{amsmath,amsfonts,bm,graphicx,dsfont}

\newcommand{\AppI}{\textrm{AprxI} }
\newcommand{\AppII}{\textrm{AprxII} }
\newcommand{\Tmat}{\textrm{Tmat} }
\newcommand{\G}{\Gamma}

\newcommand{\rmd}{\mathrm{d}}
\newcommand{\eps}{\epsilon}

\newcommand{\s}{\sigma}

\newcommand{\ixe}{\rangle\!\langle}
\newcommand{\phd}{\phantom{\dagger}}

\renewcommand{\a}{\alpha}

\newcommand{\bra}[1]{\langle #1|}
\newcommand{\ket}[1]{|#1\rangle}

\newcommand{\ua}{\uparrow}
\newcommand{\da}{\downarrow}

\usepackage{color}

\begin{document}

\title{Inelastic cotunneling in quantum dots and molecules with weakly broken degeneracies}
\author{Georg Begemann}
\author{Sonja Koller}
\author{Milena Grifoni}
\affiliation{Theoretische Physik, Universit\"at Regensburg, 93040 Regensburg, Germany}
\author{Jens Paaske}
\affiliation{The Niels Bohr Institute and Nano-Science Center, University of Copenhagen, Universitetsparken 5, DK-2100 Copenhagen {\O}, Denmark}
\date{\today}

 \begin{abstract}
We calculate the nonlinear cotunneling conductance through
interacting quantum dot systems in the deep Coulomb blockade regime
using a rate equation approach based on the $T$-matrix formalism, which shows in the concerned regions very good agreement with a generalized master equation approach.
Our focus is on inelastic cotunneling in systems with weakly broken
degeneracies, such as complex quantum dots or molecules. We find for
these systems a characteristic gate dependence of the non-equilibrium
cotunneling conductance. While on one side of a Coulomb diamond the
conductance \textit{decreases} after the inelastic cotunneling
threshold towards its saturation value, on the other side it
\textit{increases} monotonously even after the threshold. We show
that this behavior originates from an asymmetric gate voltage dependence of the effective cotunneling amplitudes.
 \end{abstract}

\pacs{85.35.Be, 73.63.kv}

\maketitle
\section{Introduction}
Quantum dot devices, or so-called artificial atoms, consist of a
small electronic nanostructure tunnel coupled to source and drain
leads. In the Coulomb-blockade regime, sequential (one-electron)
tunneling transport is exponentially suppressed and processes where
two or more electrons tunnel simultaneously become the dominant
transport mechanism.~\cite{Houten92} Among such correlated tunneling
processes, \textit{cotunneling} has received a lot of interest in
recent years. Cotunneling denotes a two-electron tunneling process
which can transfer an electron coherently from source to drain by a
virtual population of an energetically forbidden charge state of the
nanostructure. As energy is gained from the voltage drop during the
electron transfer, a cotunneling event can leave the structure in an
excited state, in which case one speaks of \textit{inelastic
cotunneling}. Otherwise the energy state of the island is left
unchanged and the process is called \textit{elastic}.

Inelastic cotunneling spectroscopy has turned out to be a useful
tool to identify electronic, magnetic and vibrational excitations in
semiconducting~\cite{Schleser05,De-Franceschi01} or carbon nanotube
based~\cite{Sapmaz05,Babic04,Paaske06,Holm08} quantum dots as well as
in single-molecule
junctions.~\cite{Parks07,Roch08,Osorio07,Osorio08,Osorio10} Most
importantly, the positions of conductance peaks provide a very
direct fingerprint of the excitation spectrum of the tunnel-coupled
nanostructure, but also the more detailed bias-dependence or {\it
line-shape} of such inelastic cotunneling peaks contains valuable
information. By now, it is well
understood~\cite{Wegewijs01,Parcollet02,Paaske04,Golovach04} how the
non-equilibrium pumping of excited states by the applied bias voltage
can give rise to a cusp in the region where the bias voltage matches the
relevant excitation energy. This effect is maximal for a symmetric setup.
Having very different tunnel couplings
to respectively source and drain electrodes implies that the
nanostructure (dot or molecule) is almost equilibrated with one
electrode and this effect no longer shows up. This was confirmed by
an experiment by Parks et al. where the opening and closing of a
mechanical break junction holding a $C_{60}$ molecule was shown to
correlate with the weakening and strengthening of such
non-equilibrium cusps near the threshold for excitation of a
vibrational mode in the system.~\cite{Parks07}

\begin{figure}[b]
\centering
\includegraphics[width=0.4\textwidth]{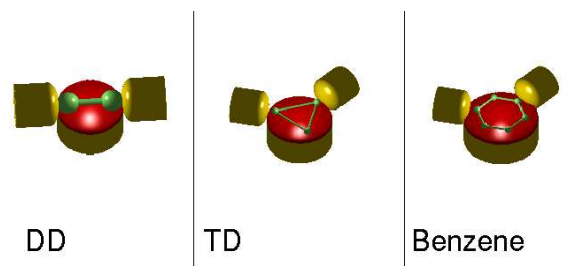}
\caption{Sketch of the setups for different quantum dot systems investigated in this work:  a double dot (DD), a triangular dot (TD), and a ``benzene'' quantum dot.}
\label{fig:setup}
\end{figure}
In the present paper we investigate the non-equilibrium cotunneling in
a variety of complex quantum dot systems. For systems designed to have low-energy
excitations arising from weakly broken degeneracies we find a characteristic gate
dependence of the non-equilibrium modulation of the inelastic cotunneling steps. On one
side of the Coulomb diamond, we find the characteristic cusped
increase in conductance at threshold, but on the other side of the
diamond this turns into a weakening of the conductance at threshold
which renders the nonlinear conductance entirely monotonous in bias.
We show how this comes about by an asymmetry in the cotunneling amplitudes
for processes which add or remove one electron from the dot ("particle-hole asymmetry"). 
This adds an important piece of information to the
spectroscopic toolbox insofar as such a non-equilibrium depression of
the cotunneling step in a symmetrically coupled device should not be
mistaken for a tunnel broadened or thermally smeared cotunneling
step in a very asymmetrically coupled system: such causes would lead to depression of the cotunneling steps at \emph{both} sides of the Coulomb blockade diamond. In contrast, the pecularity of our intrinsic effect is that a depression will occur only on one side, while on the other the conductance will retain the typical cusped increase.

The systems we study are all coupled \textit{symmetrically} to source and
drain so as to maximize the non-equilibrium effects under scrutiny.
They are chosen according to increasing complexity, namely a lateral
double-dot (DD), a triangular triple-dot (TD) and a benzene
molecule. In all cases only two dots (or sites in the benzene) are
tunnel coupled to the leads, see Figure \ref{fig:setup}. For the triple dot as well as for benzene, this induces not only a breaking of the symmetry under rotations by $n\cdot120^\circ$ respectively $n\cdot60^\circ$ ($n\in\mathbb{Z}$), but also a degeneracy lifting: While the on-site energies of all uncoupled sites can be normalized to zero, the contact sites have to be endowed with a different on-site energy $\xi$ to mimic the symmetry
breaking which is likely to result from either tunneling
renormalization~\cite{Holm08} or electrostatic
effects~\cite{Osorio07,Kaasbjerg08}.

To calculate the current and other observables in transport through
quantum dots in the Coulomb blockade regime there are a number of
techniques, each with their advantages and limitations. Following
a real-time transport approach as described in Refs.~\cite{Schoeller94,Konig96,Konig96b,Leijnse08}, one can trace out the leads degrees of freedom to
derive a formally exact generalized master equation (GME) for the
reduced density matrix (RDM) of the system. This approach allows for
a systematic expansion in the tunneling Hamiltonian $H_{\textrm T}$,
thus capturing sequential tunneling from contributions of second
order in $H_{\textrm T}$ and cotunneling as a fourth order process.
The GME has the advantage that it is exact to a desired order.
However, already at fourth order in $H_{\textrm T}$ the number of
terms is quite large and therefore it is often practical to use
simpler approaches which capture only the most relevant
contributions for each order of the perturbation theory.


In this work, we focus entirely on the Coulomb blockade regime and
we shall therefore calculate cotunneling rates within the $T$-matrix
approach.~\cite{Bruus04} This approach is much simpler than the GME,
and will be valid {\it deep inside} a Coulomb diamond, even with the
further approximations of {\it i)} neglecting the sequential
tunneling contributions, and {\it ii)} approximating the
denominators in the rates as independent of the lead electron
energy. 
To ensure that all details in the {\it lineshapes} for which we aim 
are correct, we benchmark the above mentioned approximations as well
as the $T$-matrix technique itself by a quantitative comparison to the GME approach.
In general, we find the $T$-matrix and the
GME approach to be in good agreement when additional effects due to
level shifts and broadening are irrelevant, namely in the regime
where the tunneling induced level broadening is much smaller than
the temperature.

The paper is organized as follows: In Section \ref{sec:model}, we
introduce the model Hamiltonian and the relevant expression to
calculate the current and conductance in terms of transition rates
is provided. In the end of this section, we discuss different
approximations to the rates. These approximation schemes are
compared against exact fourth order results in Section
\ref{sec:approaches} for the case of a double-dot model. The triple
dot and the benzene molecule are investigated in Section
\ref{sec:WBD} and the results are analyzed analytically using the
simplest of these approximations. Conclusions are found in Section
\ref{sec:conclusions}.

\section{The Model}\label{sec:model}
For convenience the conventions $e=1$ and $\hbar=1$ will be used throughout the paper.

A generic quantum dot coupled to source and drain leads is described
by the Hamiltonian
\begin{equation}
H = H_{\rm QD} + H_{\rm leads} + H_{\rm T}.
\end{equation}
Source and drain leads  are represented by two reservoirs of
non-interacting electrons: $H_{\rm leads} =
\sum_{\alpha\,k\,\s}(\eps_k - \mu_{\alpha})
 c^{\dagger}_{\alpha k \s}c^{\phd}_{\alpha k \s}$,
where $\alpha = L,R$ stands for the left or right lead. The chemical
potentials $\mu_{\alpha}$ of the leads depend on the applied bias
voltage $V_{b}$, which is assumed to be applied symmetrically
across the two junctions, so that $\mu_{\rm L,R} = \mu_0 \pm
\tfrac{V_{b}}{2}$. In the following we will measure the energy
starting from the equilibrium chemical potential $\mu_0 = 0$. The
Hamiltonian of the quantum dot itself depends of course on the
underlying nanostructure, be they quantum dots defined on
semiconducting heterostructures,~\cite{Kouwenhoven01} carbon
nanotubes~\cite{Bezryadin98} or molecules bridging two
contacts~\cite{Reed97}. We consider here some archetypal model for
quantum dots. They can be described as two or more ($M$) localized
states coupled among each other. Following the semi-empirical
modeling of benzene,~\cite{Pariser53,Pople53,Hettler03,Begemann08}
we study here model Hamiltonians given by
\begin{eqnarray}
\nonumber
H_{\textrm{QD}}&=& \sum^M_{j=1}\sum_{\s}\left(\epsilon_j-e\kappa V_g\right)  d^{\dagger}_{j\s}d^{\phd}_{j\s} +U\sum^{M}_{j=1}n_{j\ua}n_{j\da} \\ \label{eq:DD}
& &+ \sum_{<ij>}  b_{ij} d^{\dagger}_{i\s}d^{\phd}_{j\s} +  \sum_{\sigma\sigma'}\sum_{i<j} V_{ij}n_{i\sigma}n_{j\sigma'},
\label{eq:MQD}
\end{eqnarray}
where $\epsilon_j$ is the on-site energy of the level $j$. In Section \ref{sec:WBD} we will set $\epsilon_j=0$, except for the two sites coupled to the contacts, for which we assume $\epsilon_j=\xi\ll U,V_{ij}$, establishing a weakly broken degeneracy. The parameter $b_{ij}<0$ 
describes the hopping of electrons between nearest neighboring
states $i,j$. 
 $U$  accounts for the on-site charging energy, while
$V_{ij}$ is the interaction between two states. The first term
accounts for the influence of a gate voltage $V_g$, with $\kappa$
being the gate coupling parameter. Its actual value strongly depends on the fabrication technique used for creating the quantum dot,~\cite{Osorio08} and typically ranges in order of magnitude from $10^{-3}$-$1$. As it simply acts as a scaling factor on the gate voltage, we can set $\kappa=1$ throughout
this paper without loss of generality. The leads couple only to some of these localized states
and the corresponding tunneling Hamiltonian is described by
%
\begin{equation}
 H_{\rm T} = \sum_{ k \a \s}
 \left({t^{\alpha}}^{*} d^{\dagger}_{j_{\alpha} \s} c^{\phd}_{\alpha k \s} +
       t^{\alpha} c^{\dagger}_{\alpha k \s} d^{\phd}_{j_{\alpha} \s}\right),
\end{equation}
where $d^{\dagger}_{j_{\alpha} \sigma}$ creates an electron in the single particle state $\ket{j\sigma}$ which couples to lead $\alpha$. The tunneling Hamiltonian $H_{\textrm T}$ is treated as a perturbation to $H_{\textrm{QD}}+H_{\textrm{leads}}$.
In this work we investigate a double dot (DD), a triple dot (TD) and a benzene molecule, corresponding to $M=2,3,6$ in Eq.~(\ref{eq:MQD}), connected to source and drain as illustrated schematically in Figure~\ref{fig:setup}. We give an overview of their relevant properties in Sections \ref{sec:approaches} and \ref{sec:WBD}. As we shall show, weakly broken degeneracies in the TD and benzene give rise to qualitatively different inelastic cotunneling profiles at different gate-voltages. This is in contrast to the DD which has a non-degenerate ground state and therefore less structure in its cotunneling amplitudes.

\subsection{Calculating the cotunneling current}
To make calculations easier, we shall assume $\G\ll k_BT\ll\xi\ll E_C$. Here, $E_C$ is the addition energy, which can in principle be expressed in terms of $U$ and $V_{ij}$, but with increasing number of sites in an increasingly unhandy way. The value of the level broadening $\Gamma$ must stay well below the thermal energy $k_BT$ in order to justify a perturbative approach to transport. The degeneracy lifting $\xi$ is much smaller than the addition energy $E_C$, but exceeds both $\Gamma$ and the thermal energy by far. In this case, the reduced density matrix remains diagonal, because one can exclude coherences between the no longer degenerate states, and the rate equations are simpler.~\cite{Leijnse08,Darau09,Koerting08} Moreover, normal metal leads are considered, such that a rate equation approach to transport is sufficient.

For the sequential tunneling rates, a GME approach~\cite{Schoeller94, Konig96, Konig96b} yields the same result as Fermi's golden rule with $H_{\rm T}$ being the perturbation.
The latter scheme can be iterated to include higher order tunneling processes by making use of the $T$-matrix
\begin{equation}
\label{eq:TM}
 T(E) = H_{\textrm T}+H_{\textrm T}\frac{1}{E-H_0+i0^{+}}T.
\end{equation}
from which transition rates from the initial to the final state can be calculated up to a given order in $H_{\textrm T}$.\\
In general, the emerging $T$-matrix rates differ from the corresponding GME rates. The latter are exact to a given order in perturbation theory, and explicitly exclude all reducible terms,~\cite{Koch06,Averin94,Turek02} i.e. divergences caused by the denominator in Eq. (\ref{eq:TM}) going to zero. By construction the $T$-matrix misses in each order contributions guaranteeing these exclusion via a cancellation of reducible terms. Therefore unavoidable divergences emerge with the $T$-matrix technique from fourth order in the perturbation and onwards.~\cite{Timm08,Koller09b} Meanwhile, regularization schemes to remove the divergence appearing in the fourth order $T$-matrix rates have become standard,~\cite{Averin94, Turek02, Koch06} and the $T$-matrix approach has been applied to various setups, e.g. to a double dot structure~\cite{Golovach04} or to molecular systems where electronic and vibronic degrees of freedom can be strongly coupled~\cite{Koch06}. In this context, it is important to stress that the standard way of regularizing the $T$-matrix rates does not exactly reproduce the GME intrinsic regularization, but the discrepancy between $T$-matrix and exact perturbation theory turns out to vanish deep inside the Coulomb blockade.
The same holds for further fourth order contributions included by the GME which cannot necessarily be brought into the form of a squared matrix element~\cite{Koller09b} and are disregarded by the $T$-matrix approach.\\

We label now the states of the quantum dot with their particle number $N$, the $S_z$-component of their spin with $\eta$ and an additional quantum number with $l$. The $T$-matrix rate for a transition between two states $\ket{N' l' \eta'} \rightarrow \ket{N l \eta } $ of the quantum dot system is then given by~\cite{Bruus04}
\begin{widetext}
\begin{eqnarray}
\label{eq:gamma}
\G_{\vert N l \eta \ixe N' l' \eta' \vert} &=& 2\pi \sum_{f,i}\! \left\vert \bra{f_{N l \eta}} H_{\rm T}+  H_{\rm T}\frac{1}{E_{i_{N'l'\eta'}}-H_{\rm QD}-H_{\rm leads}+i0^{+}} {H_{\rm T}}\ket{i_{N' l' \eta'}}\right\vert^2
 W_{i_{N' l' \eta'}}\delta(E_{f_{N l\eta}}-E_{i_{N' l' \eta'}}).
\end{eqnarray}
\end{widetext}
Here the sum is over all possible initial $(i)$ and final $(f)$ states of the overall system including the leads, $\ket{i_{N' l'\eta'}} =\ket{N'l'\eta'} \ket{i_L} \ket{i_R}$, weighted by a thermal distribution function $ W_{i_{N' l' \eta'}}$. The rate from Eq. (\ref{eq:gamma}) comprises the dominant fourth order contributions deep inside the Coulomb diamonds, namely the cotunneling effects.\\

The rate equation describing the dynamics of the occupation probabilities of the states reads
\begin{eqnarray}
\label{eq:RE}
\lefteqn{\dot P^{N l \eta} =}\\ \nonumber
& &-\sum_{N' l' \eta'} \G_{\vert N' l' \eta' \ixe N l \eta \vert}\ P^{N l \eta} + \sum_{N' l' \eta'} \G_{\vert N l \eta \ixe N' l' \eta' \vert}\ P^{ N' l' \eta'},
\end{eqnarray}
where $P^{N l \eta}(t)$ is the probability of finding the dot in the state $\ket{N l \eta}$ at time $t$. The stationary solution ($t\rightarrow \infty$) is therefore given by
\begin{equation}
\label{eq:rateeq}
\sum_{N' l' \eta'} \G_{\vert N' l' \eta' \ixe N l \eta \vert} P^{N l \eta}_{\rm stat} = \sum_{N' l' \eta'} \G_{\vert N l \eta \ixe N' l' \eta' \vert}\ P^{ N' l' \eta'}_{\rm stat},
\end{equation}
with the normalization condition
\begin{equation}
\label{eq:norm}
\sum_{N l \eta} P^{N l \eta}_{\rm stat} = 1.
\end{equation}
With the help of the stationary solution, we arrive at an approximate expression for the current up to fourth order in $H_{\rm T}$,
\begin{equation}
\label{eq:current}
I = I_{\rm sequential}+I_{\rm cotunneling},
\end{equation}
with the second order,
\begin{eqnarray}
\lefteqn{
I_{\rm sequential} = } \\ \nonumber
& &\sum_{N l \eta } \sum_{l' \eta'}\left(\G^{L}_{\vert N+1 l' \eta'\ixe N l \eta \vert} -\G^{L}_{\vert N-1 l' \eta'\ixe N l \eta \vert}\right)P^{N l \eta}_{\rm stat}\label{I2}
\end{eqnarray}
and fourth order contribution
\begin{eqnarray}
\lefteqn{I_{\rm cotunneling} = }\\ \nonumber
& &\sum_{N l \eta}\left[ \sum_{ l'\eta'}\left(  \G^{RL}_{\vert N l'\eta' \ixe N l\eta \vert} - \G^{LR}_{\vert N l'\eta' \ixe N l\eta \vert}\right)\right] P^{N l \eta}_{\rm stat}.
\end{eqnarray}
Here, the superscripts to the rates indicate at which lead $\alpha$ the tunneling processes take place. Truncating the general expression (\ref{eq:gamma}) for the rate to second order, we retain only one tunneling event, which can either involve the left or the right lead. For the stationary current flow, we merely need to consider the balance between in- and out-tunneling at one of the electrodes, and we have chosen in Eq. (\ref{I2}) the left one, $\alpha=L$.

The fourth order cotunneling events transfer an electron fully across the quantum dot, which involves two tunneling events at distinct leads. Therefore the resulting current is given by the balance between charge transfer from left to right ($RL$) and from right to left ($LR$).
The cotunneling rates emerging from Eq.~(\ref{eq:gamma}) can be written as
\begin{multline}
\label{eq:geff}
\G^{\rm eff}_{\vert N l \eta \ixe N l' \eta' \vert} = 2\pi \sum_{f,i}\! \left\vert \bra{f_{N l \eta}} H^N_{\rm int} \ket{i_{N l' \eta'}}\right\vert^2\\\times
 W_{i_{N l' \eta'}}\delta(E_{f_{N l\eta}}-E_{i_{N l' \eta'}}),
\end{multline}

where $H^N_{\rm int}$ is given by
\begin{multline}\label{eq:cotunham}
H^N_{\rm int} = \sum_{\alpha k\s} \sum_{\alpha' k' \s'}t^{\a}t^{\a'} h^{l l'}_{\eta\eta'} \ket{N l\eta}\!\bra{N l'\eta'}  c^{\dagger}_{\a k\s} c^{\phd}_{\a' k'\s'},
\end{multline}
with matrix elements
\begin{multline}
h^{l l'}_{\eta\eta'}= \\
\left[\sum_{l''\eta''}\frac{ \bra{N l\eta}d_{j_{\a}\s} \ket{N+1 l''\eta''}\!\bra{N+1 l''\eta''} d^{\dagger}_{j_{\a'}\s'}\ket{N l'\eta'} }{ E_{N l'\eta'} - E_{N+1 l''\eta''} +\eps_{k'\a'}+i0^{+}}\right.\\
+ \left. \sum_{l''\eta''}\frac{\bra{N l\eta} d^{\dagger}_{j_{\a}\s} \ket{N-1 l''\eta''}\!\bra{N-1 l''\eta''} d^{\phantom{\dagger}}_{j_{\a'}\s'}\ket{N l'\eta'} }{ E_{N-1 l''\eta''} - E_{Nl'\eta'} +\eps_{k\a}-i0^{+}}\right].
\label{eq:hint}
\end{multline}
Note that the effective cotunneling Hamiltonian (~\ref{eq:cotunham}) now takes the form of a generalized Kondo, or Coqblin-Schrieffer model,~\cite{Hewson93} depending on the symmetries of the states $|N l \eta\rangle$.

\subsubsection{Approximation I}

To calculate the rates (\ref{eq:geff}) analytically, we neglect in a first approximation the $\eps_{k\alpha}$ energy dependence in the denominators of $H^N_{\rm int}$. This is justified for small (compared to the charging energy) bias voltages, so that the electrons that tunnel to and from the leads have energies around the equilibrium chemical potential and thus $|E_{N\pm1l''\eta''} -E_{N l'\eta'}|\gg |\eps_{k\alpha}|, |\eps_{k'\alpha'}|$. Converting the sums over $k,k'$ into integrals assuming a flat band with constant density of states, a simple integration leads to the cotunneling rates
%
\begin{multline}
\label{eq:LRrateI}
\G^{\rm{eff},RL}_{\ket{N\eta l}\!\bra{N\eta' l'}}= 2\pi\sum_{\s\s'} \nu_L \nu_R (-(E_{N l'\eta'}-E_{N l\eta})-V_{b})\\\times
\vert \sum_{\a\a'} \delta_{\a' L} \delta_{\a R} t^{\a}t^{\a'} h^{l l'}_{\eta\eta'}\vert^2   n_B\left(-(E_{N l'\eta'}-E_{N l\eta})-V_{b}\right),
\end{multline}
%
where $n_{B}(x)=\frac{1}{\exp(\beta x)-1}$ is the Bose function, $\beta$ is the inverse temperature and $\nu_{\a}$ is the density of states in lead $\a$.
This approximation is valid for gate and bias  voltages inside the $N$-electron Coulomb diamond. We refer to this approximation as {\bf \AppI}.
\subsubsection{Approximation II}

Alternatively, to get a more precise description of the inelastic cotunneling conductance (when $\mu_L-\mu_R > E_{Nl'} -E_{N0}$), we can take into account the energy dependence of $H_{\rm int}^N$. By shifting the integration variable $\eps_{k\alpha}\rightarrow\eps_{k\alpha}+\mu_{\alpha}$ in Eq. (\ref{eq:hint}), we see that $H_{\rm int}^N$ now explicitly depends on $\mu_{L},\mu_{R}$ and therefore on the bias voltage. In the rates, we get expressions of the form
\begin{eqnarray}
\nonumber
 \G& \sim& \int d\eps f(\eps)\left(1-f(\eps+\mu_L-\mu_R+E_{Nl\eta}-E_{Nl'\eta'})\right)\\
\label{eq:regularization}
& &\times\frac{1}{\eps-E_1\pm i0^{+}}\frac{1}{\eps-E_2\pm i0^{+}},
\end{eqnarray}
where $E_1$ and $E_2$  depend on $l\eta$ and $l'\eta'$ and the summation indices in $h^{ll'}_{\eta\eta'} $. If $E_1=E_2$, Eq.~(\ref{eq:regularization}) cannot be evaluated directly, because of divergences stemming from second order poles. This problem was stated already in 1994,~\cite{Averin94}  and a regularization scheme has been developed and become standard within the $T$-matrix approach to transport.~\cite{Turek02, Koch06} In this regularization scheme, a finite width $\gamma\sim\Gamma$ is attributed to the states which enter the denominators as imaginary parts. This level broadening physically stems from the tunnel coupling, but is not taken into account by the $T$-matrix approach.  Thus the poles are shifted away from the real axis so that the integral can actually be performed. The resulting expression can be expanded in powers of $\gamma$ and the leading term is found to be of order $1/\gamma$. Together with the prefactor of the rates, $\Gamma^2$, this term is identified to be a sequential tunneling term. It is excluded to avoid double counting of sequential tunneling processes. The next to leading order term is of order $\gamma^0$ and gives the regularized cotunneling rate. At this point, the actual value of the broadening does not matter and the limit $\gamma\rightarrow0$ can safely be taken. The calculation of the current with regularized cotunneling processes and disregarding sequential tunneling rates {($I_{\rm sequential}=0$ in Eq. (\ref{eq:current}))}, we  refer to as {\bf \AppII}.

\subsubsection{$T$-matrix}
Both \AppI and \AppII are expected to fail when cotunneling assisted sequential tunneling processes become accessible.~\cite{Golovach04,Schleser05,Huttel09} This can happen well inside the Coulomb diamond, when excited $N$ particle states are populated via inelastic cotunneling. Indeed, at the lines given by the equation $\mp V_g\pm E_{N\pm1\eta}\mp E_{N l'\eta'} + \mu_{\a} = 0,\, l'\neq 0$ (dashed lines inside the Coulomb diamond in Fig. \ref{fig:sketch}) the cotunneling rates become negative which leads to an ill-defined set of rate equations, unless we include also sequential tunneling terms and allow also states with $N\pm1$ to be populated. This is exactly the $T$-matrix approach, referred to as {\bf \Tmat} in the following.
\section{Inelastic cotunneling IN a double dot}\label{sec:approaches}
In this section we discuss inelastic cotunneling features of the simplest model described by Eq.~(\ref{eq:DD}), namely by a double dot system (DD).
Additionally, it is used as a benchmark for the $T$-matrix approach Tmat as well as for \AppI and \AppII against a calculation based on the GME approach.
\begin{figure}
\centering
\includegraphics[width=0.49\textwidth,angle=0]{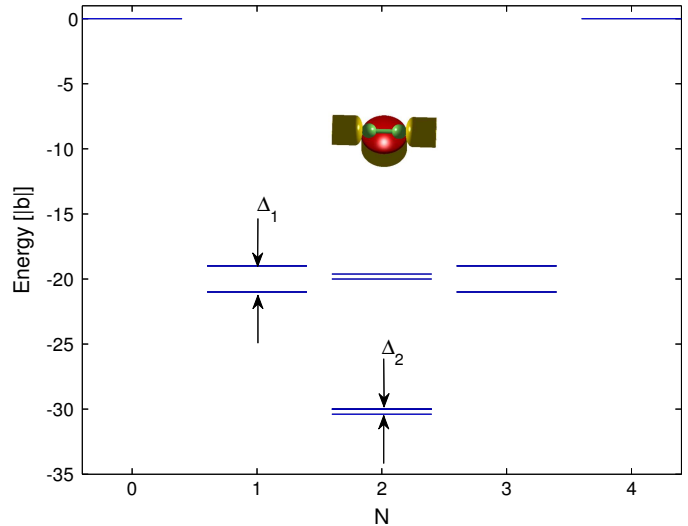}
\caption{Spectrum of a double quantum dot, as described by Eq.~(\ref{eq:DD}) with $M=2$. The gate voltage was set to $V_g = 20|b|$, which corresponds to the center of the $N=2$ diamond (see also Figures~\ref{fig:sketch} and \ref{fig:log10dIdV}). The splitting of the $N=1$ states is $\Delta_1=-2b$, the singlet-triplet splitting for the $N=2$ states is $\Delta_2=0.5(V-U+\sqrt{16b^2+(U-V)^2})$. The interaction parameters are $U = 20|b|$, $V = 10|b|$.}
\label{fig:specDD}
\end{figure}
\begin{figure}
\centering
\includegraphics[width=0.49\textwidth,angle=0]{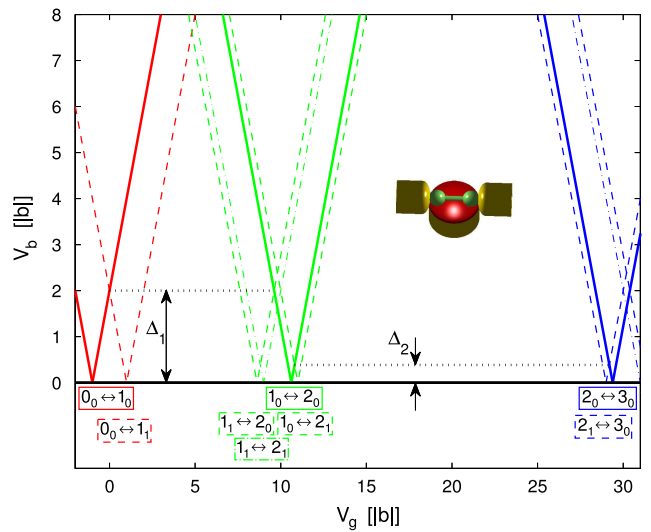}
\caption{Sketch of the stability diagram for the double quantum dot (DD) described by Eq.~(\ref{eq:DD}) with $M=2$ and parameters $U = 20 |b|$, $V = 10|b|$. Red lines indicate a transition between states with zero and one, green lines between states with one and two, blue lines between two and three electrons. Solid lines are for ground state to ground state transitions and define the Coulomb blockade regions, dashed lines involve excited states. We have labelled the participating states by $N_i$. It indicates the $i$th $N$ electron state, with associated energy $E_i$ (for example, the one electron ground state is labeled with $1_0$, the first excitation with $1_1$, and so on). Furthermore, the dotted lines indicate the onset of the inelastic cotunneling at $V_b=\Delta_1$ in the $N=1$ and at $V_b=\Delta_2$ in the  $N=2$ diamond.}
\label{fig:sketch}
\end{figure}
\begin{figure}
\centering
\includegraphics[width=0.49\textwidth,angle=0]{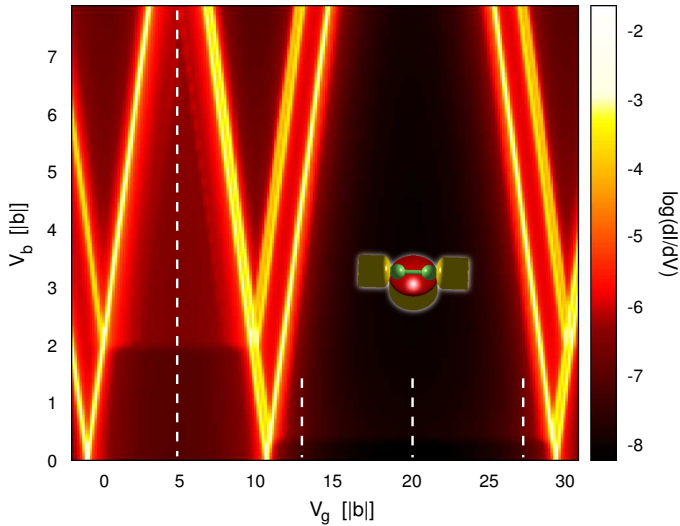}
\caption{Logarithm ($\log_{10}$) of the differential conductance $\d I/\d V_b$ for the double quantum dot (DD) as a function of gate and bias voltage  calculated with the $T$-matrix approach. One recognizes the features in the $dI/dV$ discussed schematically in Figure~\ref{fig:sketch}. Vertical dashed lines indicate the cuts used in Figures~\ref{fig:cutN1} and \ref{fig:cutsN2}. Parameters are as in Figures~\ref{fig:specDD},~\ref{fig:sketch}, together with $k_{\rm B}T=0.02|b|$, $\G^{L}=\G^{R}=0.008 |b|$. Here, as well as in all following plots, the differential conductance is measured in units of $\Phi\times e^2/h$, with the scaling factor $\Phi:=\Gamma^L\Gamma^R/|b|^2$.}
\label{fig:log10dIdV}
\end{figure}
\begin{figure}
\centering
\includegraphics[width=0.49\textwidth,angle=0]{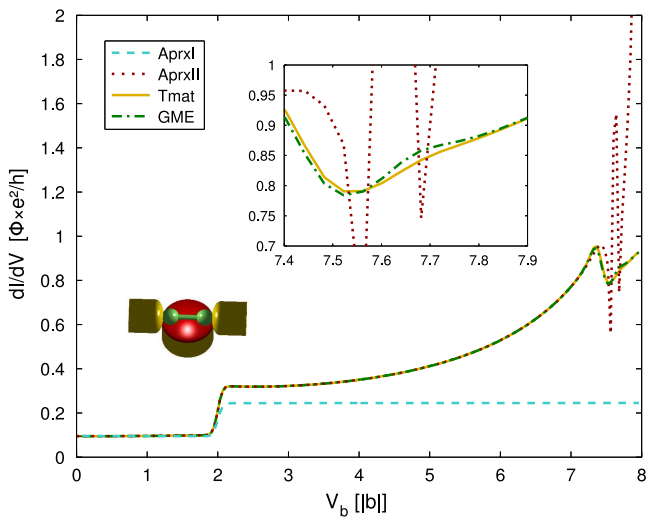}
\caption{Differential conductance $\d I/\d V_b$ as a function of bias voltage calculated with the different approximation schemes discussed in Section \ref{sec:model} as well as with the {GME} at the center of the $N=1$ diamond corresponding to $V_g=4.8 |b|$. \AppII yields divergences in the conductance at resonances. The {\Tmat and GME} show features at these positions but are well behaved. They agree almost exactly. Parameters are as in Figure~\ref{fig:log10dIdV}.}
\label{fig:cutN1}
\end{figure}
\begin{figure}
\centering
\includegraphics[width=0.49\textwidth,angle=0]{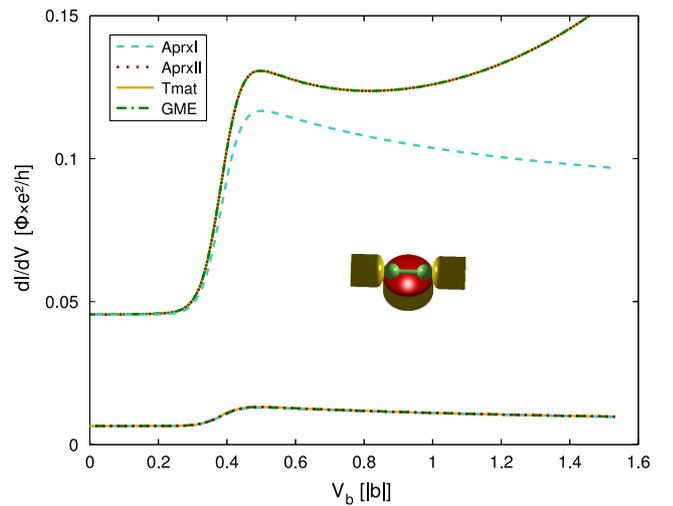}
\caption{Differential conductance $\d I/\d V_b$ as a function of bias voltage calculated with different approximation schemes in the $N=2$ diamond at $V_g=13|b|$ (upper set of lines) and at the center of the diamond corresponding to $V_g=20|b|$ (lower set of lines). Due to the particle-hole symmetry of the DD, the two cuts at $V_g=13|b|$ and $V_g=27|b|$ give exactly the same result.}
\label{fig:cutsN2}
\end{figure}

The spectrum of the DD system is shown in Figure~\ref{fig:specDD} at $V_g = 20|b|$, corresponding to the center of the $N=2$ diamond in Figures \ref{fig:sketch}, \ref{fig:log10dIdV}. The $N=1$ states are even and odd combinations of electron states on the left and right dot with energies $E_{\rm e/o}=\pm b$. For the $N=2$ states, we have a singlet ground state and an excited triplet state, separated by $\Delta_2=0.5(V-U+\sqrt{16b^2+(U-V)^2})$. Notice the particle-hole symmetry of this system, which is responsible for the symmetry of the stability diagram around this value of the gate voltage.

In Figure \ref{fig:sketch}, we show a sketch of the stability diagram for the DD together with additional excitation lines. All the lines follow from energetical considerations involving the spectrum, see Figure~\ref{fig:specDD}, and the chemical potential of the leads. We focus on the energy range relevant to the case where the dot is singly or doubly occupied, i.e. to the Coulomb diamonds with $N=1$ and $N=2$. Red lines indicate positions for transitions between states with zero and one, green lines between states with one and two, and blue lines between states with two and three electrons, respectively. Solid lines are for ground state to ground state transitions and define the Coulomb blockade regions with $N=1$ and $N=2$, dashed lines involve excited states. The dotted horizontal lines indicate the onset of the inelastic cotunneling in the two diamonds.

In Figure \ref{fig:log10dIdV}, the conductance through the DD calculated with $\Tmat$ is plotted on a logarithmic color scale.
One can see nicely the features in the $\d I/\d V_b$ at the positions of the lines in Figure~\ref{fig:sketch} and the general resemblance of the two figures. Inside the diamonds, at $V_{b}=\Delta_1$ or $V_{b}=\Delta_2$ the threshold for inelastic cotunneling is visible as horizontal lines. The onset of the cotunneling assisted sequential tunneling (see dashed lines in Figure \ref{fig:sketch}) can be noticed best in the $N=1$ diamond. Outside the diamonds, all the sequential tunneling lines can be seen.

We are now going to compare the different approximation schemes for the cotunneling rates as discussed in the previous section with the exact perturbation theory (GME). In Figure~\ref{fig:cutN1}, we show the differential conductance as a function of the bias voltage at the center of the $N=1$ diamond, as indicated by the dashed white line in Figure~\ref{fig:log10dIdV}.
We see that \AppI yields good agreement with the {GME} only at small bias voltages. In particular, the lineshape at the inelastic cotunneling threshold is not reproduced correctly, because the condition of validity for \AppI, $\Delta \ll E_C$, is not fulfilled here. We see that the other approaches predict an increasing differential conductance for larger bias voltages, which can be understood from the $\eps_k$ dependence in the denominators in Eq.~(\ref{eq:hint}).

\AppII and {GME} agree nicely as long as gate and bias voltages are such that one is in the innermost diamond defined by the dashed lines (see Figure \ref{fig:sketch}). Outside of this region, \AppII is no longer valid: Once the cotunneling assisted sequential tunneling sets in, the cotunneling rates in \AppII can become negative and the rate equations are ill-defined.

Inside the overall Coulomb diamond, the {\Tmat and GME} yield almost exactly the same result. Small relative deviations (few per cent) between the two approaches can be seen at the resonant lines (see inset in Figure~\ref{fig:cutN1}), which can be attributed to a certain class of terms in the rates not taken into account by the $T$-matrix.~\cite{Koller09b}

In the $N=2$-particle diamond, a better separation of the energy scales defined by the addition energy and the inelastic cotunneling threshold is given. As expected, all approximation schemes and the GME give almost exactly the same result at the center of the diamond (see lower set of lines in Figure~\ref{fig:cutsN2}). More towards the $N=1,2$ charge degeneracy point, at $V_g=13|b|$, we see that \AppI gives still a good qualitative description of the lineshape of the conductance, but as in the $N=1$ diamonds it fails to reproduce the increase of the conductance due to the bias dependence of the denominators of the rates. The decrease of the conductance after the inelastic cotunneling threshold is due to the non-equilibrium redistribution of the population of the excited state. At low bias, only the ground state is populated, and only cotunneling processes that do not change the occupation of the ground state are possible. When the bias is large enough to populate the excited state, the conductance suddenly increases due to the new possibilities of transferring electrons from left to right lead. The excited state starts to acquire a finite non-equilibrium population from this point on, and together with the increasing depopulation of the ground state this leads typically to a decrease of the differential conductance after the sudden increase at the threshold. We will discuss the behavior of the conductance at the inelastic cotunneling threshold  in more detail in the next section.

Since the DD is particle-hole symmetric, cuts through the $N=2$ diamond at the same distances from the center towards the $N=1,2$ and $N=2,3$ charge degeneracy points give exactly the same result.

\section{Inelastic cotunneling in dots with weakly broken degeneracies}\label{sec:WBD}
\begin{figure}
\centering
\includegraphics[width=0.49\textwidth,angle=0]{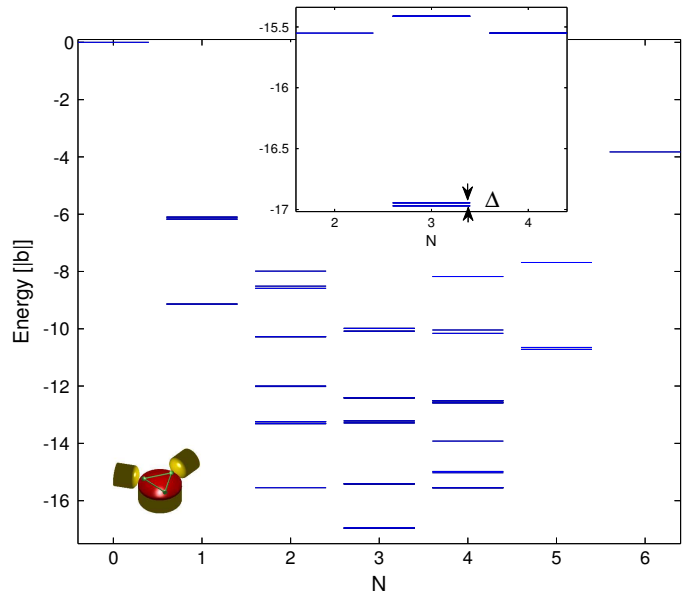}
\caption{Spectrum of the triple dot at $V_g=1.14|b|$ (center of the $N=3$ Coulomb diamond). The addition energy $E_C=U^{\rm add}_{N=3}$ for $N=3$ can be read off as the distance between the $N=3$  and the $N=2$ ground state plus the distance between the $N=4$ and the $N=3$ ground state. The splitting $\Delta$ of the $N=3$ ground state is about a hundred times smaller than the addition energy (see inset). Parameters are $U=5|b|$, $V=2|b|$, $\xi=-0.1|b|$.}
\label{fig:specTD}
\end{figure}

Systems with slightly broken symmetries, e.g. molecules in a single-molecule junction, exhibit weakly broken degeneracies, where the splitting of the originally degenerate states is much smaller than the addition energy. These systems provide a separation of energy scales which allows us to investigate inelastic cotunneling effects that are largely unaffected by charge fluctuations. From a technical point of view, this brings us into the validity range of the simplest approximation on the cotunneling rates (\AppI).\\We assume in the following a site independent hopping $b_{ij}=b<0$ for nearest neighbors $i,j$ and a shift of the on-site energies of the contacted sites by $\xi=-0.1|b|$.

\subsection{Triangular triple dot}
A triple quantum dot (TD) system is described by $H_{\rm QD}$ in Eq.~(\ref{eq:MQD}) with $M=3$. We assume that the left lead is coupled to dot 1 and the right lead to dot 2.
The coupling of dots 1 and 2 to the leads breaks the symmetry of the isolated molecule, and in accordance with our previous statements we set thus $\eps_2 = \eps_1 = \xi$, $\eps_3=0$. The corresponding energy spectrum is shown in Figure~\ref{fig:specTD} as a function of the number $N$ of electrons in the triple quantum dot. The gate voltage is chosen such that the lowest energy occurs when the TD is filled with $N=3$ electrons.

\begin{figure}[t]
\centering
\includegraphics[width=0.49\textwidth]{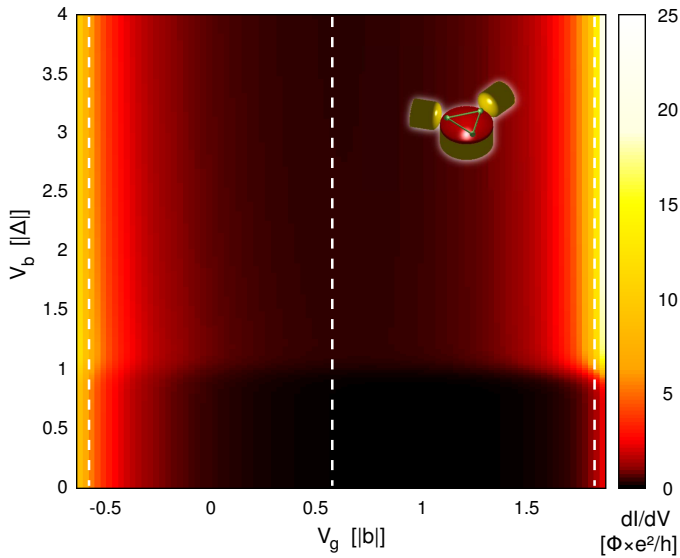}
\caption{Color coded cotunneling conductance of a TD well inside the $N=3$ Coulomb diamond. The conductance in general is lowest at the center of the diamond, and increasing when moving the gate towards the diamonds with $N\pm1$. The onset of inelastic cotunneling at $V_b=\Delta$ is clearly visible as a jump in the conductance. The charge degeneracy points are at $V_g=-1.68|b|$ (with $N=2$) and at $V_g=3.96|b|$ ($N=4$). Parameters as in Figure~\ref{fig:specTD}, with additionally $k_B T=8\times 10^{-4}|b|$, $\G^L=\G^R = 8\times 10^{-5} |b|$.}
\label{fig:diamondTD}
\end{figure}

For $\xi=0$ the $N=3$ ground state is both spin and orbitally degenerate. We label these orbitals with $l=1,2$. For finite $\xi$, they will split by an energy of $\vert E(l=2)-E(l=1)\vert=\Delta(\xi)$, where $\Delta(\xi=-0.1|b|)$ is much smaller than the addition energy $E_C=U^{\rm add}_{N=3}$ (see Figure~\ref{fig:specTD}, inset). The next excited state with $N=3$ is separated by an energy comparable to the addition energy and can thus be disregarded.

\begin{figure}
\centering
\includegraphics[width=0.49\textwidth,angle=0]{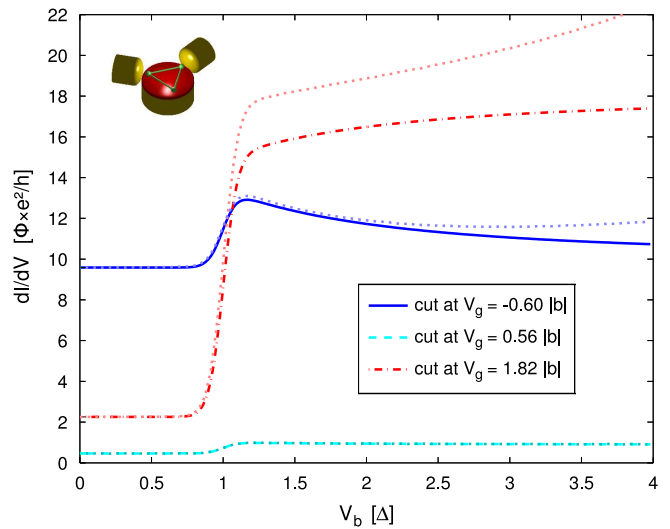}
\caption{Cotunneling conductance of a TD vs bias voltage at the positions indicated by the white dashed lines in Figure~\ref{fig:diamondTD}. The lineshape as well as the magnitude depend on the gate voltage. The solid/dashed/dot-dashed curves are calculated with \AppI, their dotted companions with \AppII. The two approaches agree well for low bias voltages $V_b\ll U^{\rm add}_{N=3}$.}
\label{fig:cutsI}
\end{figure}

In Figure \ref{fig:diamondTD}, we focus now on the situation when the system is filled with three electrons. For low temperatures, sequential tunneling is exponentially suppressed at small bias voltages, and the current is dominated by cotunneling events. We show the cotunneling conductance calculated with \AppI as a function of gate and bias voltage. The inelastic cotunneling threshold is clearly seen as a horizontal line at $V_b=\Delta$. In Figure \ref{fig:cutsI}, we show three cuts of the cotunneling conductance at different gate voltages (calculated with  \AppI and  \AppII), one at the center, two towards the corners of the $N=3$ diamond, as indicated by the dashed white lines in Figure \ref{fig:diamondTD}. As one can see from the comparison of the three cuts in Figure \ref{fig:cutsI}, the magnitude of the conductance as well as the exact lineshape now depends strongly on the gate voltage. At the center of the diamond (at $V_g\approx0.56|b|$), and even better pronounced at lower gate voltages (e.g. at $V_g=-0.60|b|$), the conductance shows the expected behavior: it is constant below the inelastic cotunneling threshold, and  above it shows a step with a cusp. The origin of this cusp lies in the non-equilibrium redistribution of the occupation probabilities of the two orbitals. For bias voltages below the threshold, only the ground state is populated. Above, the occupation probability of the excited state rises and increases with the bias, heading towards its saturation value. Until the saturation value is reached, the conductance will change with the bias. This also true for the cut at $V_g\approx 1.82 |b|$, however, there is no cusp, but a steady further increase in conductance above the inelastic cotunneling step. To understand this different behavior, we analyze the expression for the cotunneling current and the underlying rates.

We allow only the orbitals of the split ground state to be populated. We therefore have to solve the rate equations (\ref{eq:rateeq}), (\ref{eq:norm}) for $P^{N=3l\eta}$. For zero magnetic field, we expect $P^{N=3l\ua}=P^{N=3l\da}$ and conveniently we can reduce the problem to two independent variables $P_{l}=\sum_{\eta}P^{N=3l\eta},\ l=1,2$, where the index $N=3$ has been dropped. The following analysis is now performed under the assumptions that $T=0$, $V_b>0$ and $E(l=1)<E(l=2)$ without loss of generality.

The current is then given by
\begin{equation}
 I = \sum_{ll'}\Gamma^{RL}_{\ket{l'}\!\bra{l}}P_{l},
\end{equation}
and the differential conductance follows as
\begin{equation}
 \frac{\rmd I}{\rmd V_b}=\sum_{ll'}\left[\Gamma^{RL}_{\ket{l'}\!\bra{l}}\frac{\rmd P_{l}}{\rmd V_b}+P_{l}\frac{\rmd}{\rmd V_b}\Gamma^{RL}_{\ket{l'}\!\bra{l}}\right].
\end{equation}
Here, $\Gamma^{RL}_{\ket{l'}\!\bra{l}}$ is the cotunneling rate for changing the quantum dot from the state $l$ to $l'$ and thereby transferring an electron from the left to the right lead. Within \AppI, we can write the total rate $\Gamma_{\ket{l'}\!\bra{l}}$  as in (\ref{eq:LRrateI})
\begin{multline}
 \Gamma_{\ket{l'}\!\bra{l}}=\sum_{\a\a'}\Gamma^{\a\a'}_{\ket{l'}\!\bra{l}}\\
 =\sum_{\a\a'}\gamma^{\a\a'}_{\ket{l'}\!\bra{l}} \Theta\left(\mu_{\a'}\!-\! \mu_{\a}+E_{l'}-E_{l}\right) \left(\mu_{\a'}\! - \!\mu_{\a}+E_{l'}-E_{l}\right),
\end{multline}
where $\gamma^{\a\a'}_{\ket{l'}\!\bra{l}}$ depends on the gate voltage only and the remaining terms depend only on the bias voltage.
We are especially interested in the conductance slightly above the inelastic cotunneling threshold, where $V_b = \Delta+\eps, \eps\rightarrow 0^+$. At this point, still $P_1\approx1$ and $P_2\approx0$, while $\left.\frac{\rmd P_1}{\rmd V_b}\right\vert_{V_b = \Delta+\eps}<0$, $\left.\frac{\rmd P_2}{\rmd V_b}\right\vert_{V_b = \Delta+\eps}>0$ and $\frac{\rmd}{\rmd V_b}\Gamma^{RL}_{\ket{l'}\!\bra{l}} =\gamma^{RL}_{\ket{l'}\!\bra{l}}$. With these inputs, we get for the conductance
\begin{multline}
\label{eq:cond}
 \left.\frac{\rmd I}{\rmd V_b}\right\vert_{V_b = \Delta+\eps} = \left[\Gamma^{RL}_{\ket{1}\!\bra{1}} +\Gamma^{RL}_{\ket{2}\!\bra{1}}-\Gamma^{RL}_{\ket{1}\!\bra{2}}-\Gamma^{RL}_{\ket{2}\!\bra{2}}\right]\frac{\rmd P_{1}}{\rmd V_b}\\
  +\left(\gamma^{RL}_{\ket{1}\!\bra{1}} +\gamma^{RL}_{\ket{2}\!\bra{1}}\right)P_1+\left(\gamma^{RL}_{\ket{1}\!\bra{2}} +\gamma^{RL}_{\ket{2}\!\bra{2}}\right)P_2.
\end{multline}
From this expression, one sees that if $\gamma^{RL}_{\ket{2}\!\bra{2}}$ is large, the contribution containing $P_2$, which grows with raising bias, can win, so that the conductance does not show a cusp, but increases monotonously after the step. In other words, with a sufficiently large value of $\gamma^{RL}_{\ket{2}\!\bra{2}}$, the conductance will keep growing once the state $l=2$ starts to be increasingly occupied at $V_{b}\agt\Delta$. Below the threshold, the elastic cotunneling conductance is set by the prefactor to $P_1$ in Eq.~(\ref{eq:cond}). At very large bias, however, the two states become equally populated, the first line in Eq.~(\ref{eq:cond}) vanishes and the with a large value for $\gamma^{RL}_{\ket{2}\!\bra{2}}$,  the saturation conductance can become much larger than the elastic sub-threshold conductance. This implies that a monotonous increase of the conductance across the threshold will be accompanied by a large step-height, i.e. a large difference between the conductance at $V_b=0$ and at $V_b\gg\Delta$. This is clearly seen to be the case in Figure~\ref{fig:cutsI}.

\begin{figure*}[t]
\centering
\includegraphics[width=0.49\textwidth,angle=0]{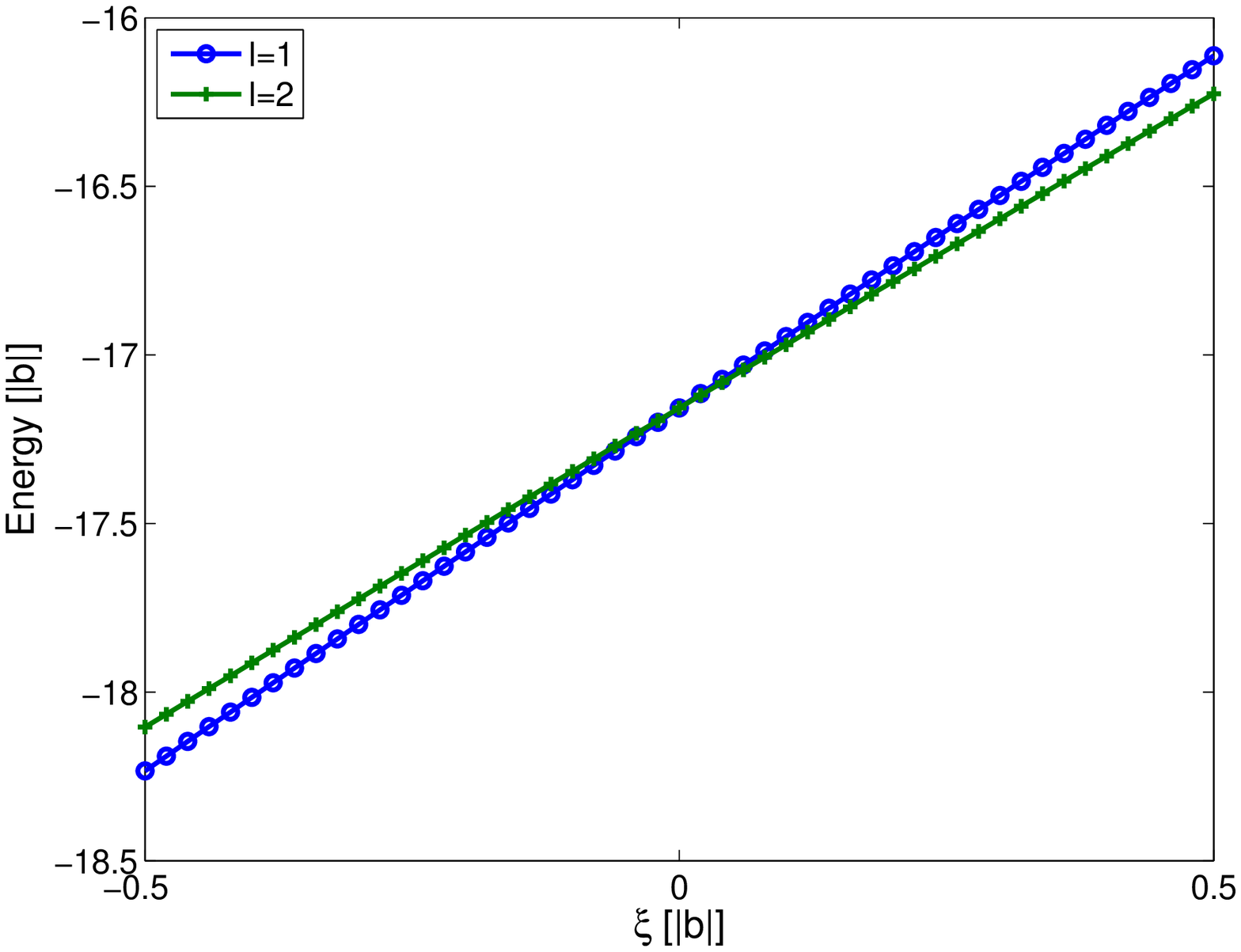}
\includegraphics[width=0.45\textwidth,angle=0]{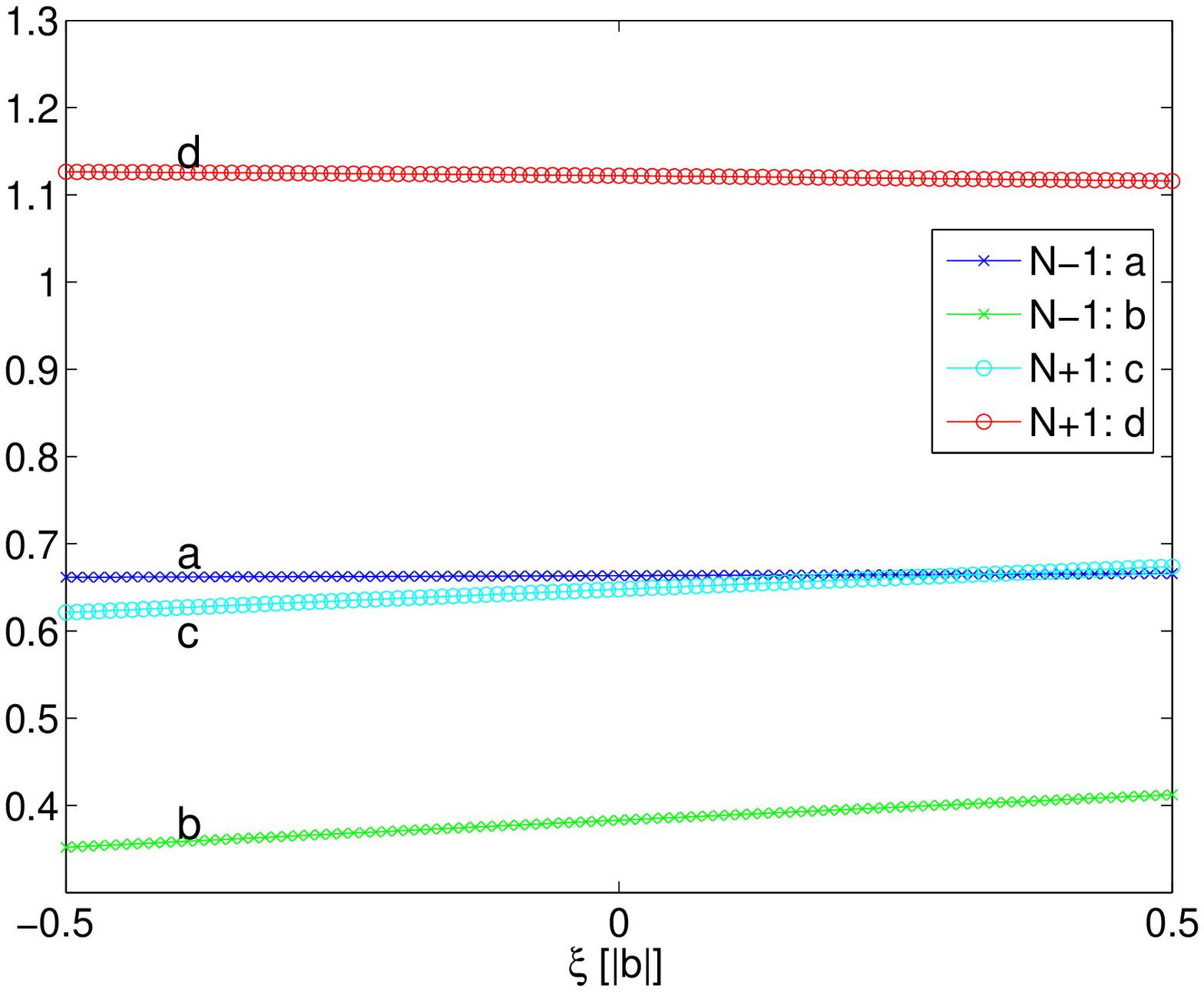}
\caption{Left panel: The quasi degenerate $N=3$ ground states of a TD with $l=1,2$ as a function of the degeneracy lifting parameter $\xi$. For $\xi<0$, the state with $l=1$ is the ground state, for $\xi>0$ the $l=2$ state has lower energy. Right panel: Overlap matrix elements of the levels $l = 1,2$  to the ground states of $N\pm1$ particles. a: $\bra{ N-1 g}\sum_{\s} d_{j_{\a}\s} \ket{N l=1}$, b: $\bra{ N-1 g}\sum_{\s} d_{j_{\a}\s} \ket{N l=2}$, c:  $\bra{N l=1} \sum_{\s}d_{j_{\a} \s}\ket{ N+1 g}$,  d: $\bra{N l=2} \sum_{\s}d_{j_{\a} \s}\ket{ N+1 g}$.}
\label{fig:levelsvsxi}
\end{figure*}
One can make the above statements about {\it large} couplings more precise, by inserting the stationary solutions for $P_1$ and $P_2$ given by
\begin{equation}
 P_1 = \frac{\Gamma_{\ket{1}\!\bra{2}}} {\Gamma_{\ket{1}\!\bra{2}}+\Gamma_{\ket{2}\!\bra{1}}},\quad P_2 = \frac{\Gamma_{\ket{2}\!\bra{1}}}{\Gamma_{\ket{1}\!\bra{2}}+\Gamma_{\ket{2}\!\bra{1}}},
\end{equation}
into the second derivative of the current at $V_b = \Delta+\eps$. From this, one obtains
\begin{widetext}
 \begin{eqnarray}
 \left.\frac{\rmd^2I}{\rmd V_b^2}\right\vert_{V_b = \Delta+\eps}&=&-2\,{\frac {\gamma^{RL}_{\ket{2}\!\bra{1}}\, \left(  \left( \gamma^{RL}_{\ket{1}\!\bra{2}}-\gamma^{RL}_{\ket{2}\!\bra{1}} \right)  \left( \gamma^{RL}_{\ket{1}\!\bra{1}}-\gamma^{RL}_{\ket{2}\!\bra{2}}
 \right)
+4\,\gamma^{RL}_{\ket{2}\!\bra{1}}\,\gamma^{RL}_{\ket{1}\!\bra{2}}
 \right) }
{\Delta\, \left( 2\,\gamma^{RL}_{\ket{1}\!\bra{2}}+\gamma^{LL}_{\ket{1}\!\bra{2}}+\gamma^{RR}_{\ket{1}\!\bra{2}} \right) ^{2}}}\\
\nonumber
& &-2\,{\frac {\gamma^{RL}_{\ket{2}\!\bra{1}}\, \left(
\gamma^{RL}_{\ket{2}\!\bra{1}}+\gamma^{RL}_{\ket{1}\!\bra{1}} -\gamma^{RL}_{\ket{1}\!\bra{2}}-\gamma^{RL}_{\ket{2}\!\bra{2}} \right)
\left( \gamma^{LL}_{\ket{1}\!\bra{2}}+\gamma^{RR}_{\ket{1}\!\bra{2}}\right) }
{\Delta\, \left( 2\,\gamma^{RL}_{\ket{1}\!\bra{2}}+\gamma^{LL}_{\ket{1}\!\bra{2}}+\gamma^{RR}_{\ket{1}\!\bra{2}} \right) ^{2}}},
\end{eqnarray}
\end{widetext}
which is positive, thus giving rise to a monotonously increasing cotunneling conductance, whenever
\begin{widetext}
\begin{equation}
 \gamma^{RL}_{\ket{2}\!\bra{2}} > \gamma^{RL}_{\ket{1}\!\bra{1}}+{\frac { \left(\gamma^{RL}_{\ket{2}\!\bra{1}} -\gamma^{RL}_{\ket{1}\!\bra{2}} \right) \left(\gamma^{LL}_{\ket{1}\!\bra{2}}+\gamma^{RR}_{\ket{1}\!\bra{2}}\right)+4\,\gamma^{RL}_{\ket{2}\!\bra{1}}\,\gamma^{RL}_{\ket{1}\!\bra{2}}} {\gamma^{RL}_{\ket{1}\!\bra{2}}-\gamma^{RL}_{\ket{2}\!\bra{1}}+\gamma^{LL}_{\ket{1}\!\bra{2}}+\gamma^{RR}_{\ket{1}\!\bra{2}}}}.
\end{equation}
\end{widetext}
The question now remains, under which circumstances this condition can be fulfilled. To answer this question, we have to analyze the transition amplitudes
\begin{widetext}
\begin{multline}
 \gamma^{RL}_{\ket{l}\!\bra{l'}}=\sum_{\eta\eta'}
\left[\sum_{l''\eta''}\frac{ \bra{N l\eta}d_{j_{R}\s} \ket{N+1 l''\eta''}\!\bra{N+1 l''\eta''} d^{\dagger}_{j_{L}\s'}\ket{N l'\eta'} }{ E_{N l'} - E_{N+1 l''} }\right.\\
+ \left. \sum_{l''\eta''}\frac{\bra{N l\eta} d^{\dagger}_{R\s} \ket{N-1 l''\eta''}\!\bra{N-1 l''\eta''} d^{\phantom{\dagger}}_{j_{L}\s'}\ket{N l'\eta'} }{ E_{N-1 l''} - E_{Nl'} }\right].
\end{multline}
\end{widetext}
We see that they depend on the overlap matrix elements of the tunneling Hamiltonian in the numerator, and on the energy differences of the states involved in the cotunneling process in the denominator. $\Delta$ is small compared to the addition energy, and we keep a distance to the edges of the diamonds, so that for our analysis, we can set $E^{N=3}(l=1)=E^{N=3}(l=2)$ in the denominator of the above expression. However, as we approach one of the two charge degeneracy points (either $N\leftrightarrow N+1$ or $N-1\leftrightarrow N$)  on the axis of the gate voltage, the contributions
\begin{equation}
 \frac{ \bra{N l} d^{\phd}_{j_{R} \s}\ket{ N+1 g}\!\bra{ N+1 g} d^{\dagger}_{j_{L}\s'} \ket{Nl'} }{ E_{Nl'} - E_{N+1g} }
\end{equation}
or
\begin{equation}
\frac{ \bra{N l}  d^{\dagger}_{j_{L}\s} \ket{ N-1g}\!\bra{ N-1 g} d^{\phd}_{j_{R}\s'} \ket{Nl'} }{ E_{N-1 g} - E_{Nl'} },
\end{equation}
with $\ket{N\pm1g}$ being the ground states with $N\pm1$ electrons,
are dominant in Eq.~(\ref{eq:gamma}). The excited states
contribute as well, but less due to the energy difference in the
denominator, and their influence will not change the qualitative
behavior of the differential conductance. It is therefore necessary
to analyze separately the matrix elements $\bra{N l}
\sum_{\s}d^{\phd}_{j_{\a} \s}\ket{ N+1 g}$ and $\bra{ N-1
g}\sum_{\s} d^{\phd}_{j_{\a}\s} \ket{Nl'}$. These are compared in
Figure~\ref{fig:levelsvsxi}, where we also investigate the behavior of the quasi degenerate levels as a
function of the degeneracy lifting $\xi$. Plotting the energies of the two levels with $l=1,2$ versus $\xi$, we see that for $\xi<0$, the state with $l=1$
is the ground state, but for $\xi>0$ the state with $l=2$ has lower
energy. The overall dependence of the matrix elements
on $\xi$ is rather weak, but the coupling to the ground states with $N\pm1$ electrons of
these two states is seen to be very different. The matrix element of the $N-1$
ground state with $l=1$ is about twice as large as the one with
$l=2$, while for the elements with the $N+1$ ground state, this
situation is reversed. This is reason why by changing the gate-voltage one can tune the system into a configuration
where $ \gamma^{RL}_{\ket{2}\!\bra{2}}$ by far exceeds
$\gamma^{RL}_{\ket{1}\!\bra{1}}$ such that the conductance
increases monotonously even after the inelastic cotunneling threshold.

The cuts in Figure \ref{fig:cutsI} were done for $\xi<0$. Choosing instead $\xi>0$, the picture would be reversed and the conductance at lowest gate voltages (close to the side of the $N-1$-diamond) would be monotonously increasing, while the conductance close to the
$N+1$-diamond would now show the {\it cusped} lineshape.
\begin{figure}
\centering
\includegraphics[width=0.49\textwidth,angle=0]{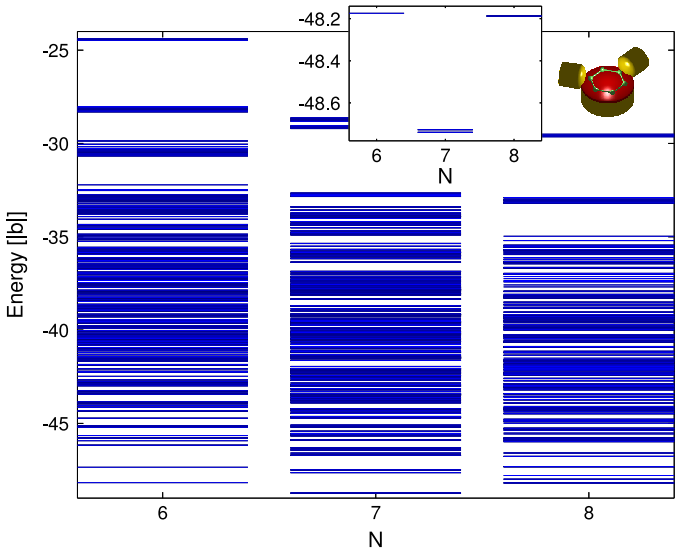}
\caption{Spectrum of the benzene molecule for $N=6$ (neutral molecule), $N=7$ and $N=8$ electrons. The initially degenerate $N=7$ ground state is split due to the coupling to the leads (see inset). Parameters are $U=4|b|$, $V=2.4|b|$,  $\xi=-0.1|b|$. }
\label{fig:spectrumBenzene}
\end{figure}
\begin{figure}
\centering
\includegraphics[width=0.49\textwidth,angle=0]{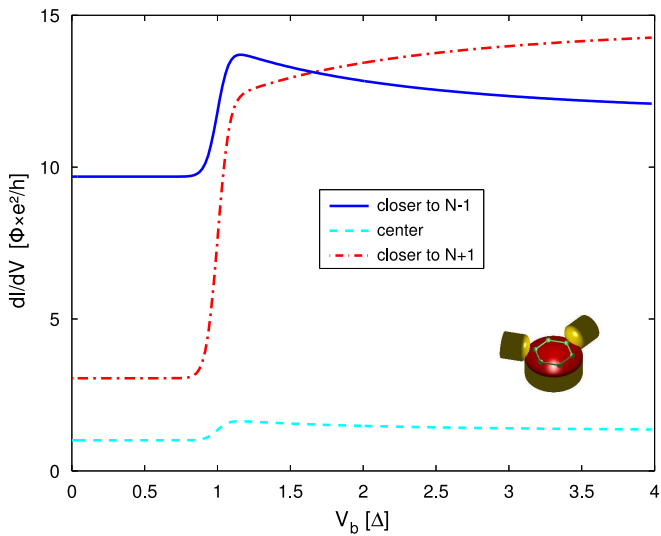}
\caption{Cotunneling conductance vs bias voltage for a benzene molecule with $N=7$ electrons coupled to the leads in meta configuration. The three curves correspond to three values of the gate voltage being closer to the $N=6$ (neutral) particle diamond, at the center of the $N=7$ diamond and closer to the $N=8$ diamond. Parameters are $U=4|b|$, $V=2.4|b|$ (we consider nearest neighbor interaction only), $\xi=-0.1|b|$, and the remaining as in Figure~\ref{fig:diamondTD}. }
\label{fig:cutsBenzene}
\end{figure}

\subsection{Benzene}

We find exactly the same effect for a singly charged ($N=7$) benzene
molecule coupled to the leads in meta configuration. The spectrum of
benzene exhibits a lot of degeneracies due to the $D_{6h}$ symmetry
of the molecule.~\cite{Begemann08} The environment of a molecular
junction can break the perfect symmetry of the molecule in various
ways.~\cite{Darau09} As for the triple-dot, we model this by
ascribing a different on-site energy to the contact sites. We
diagonalize $H_{\rm QD}$ exactly, and use the eigenstates and
eigenvalues for each charge-state to calculate all relevant
cotunneling rates. The energy-spectrum is shown in
Figure~\ref{fig:spectrumBenzene} and we now focus on the inelastic
cotunneling corresponding to the weakly broken degeneracy in the
$N=7$ state. In Figure~\ref{fig:cutsBenzene}, we show three cuts
through the $N=7$ Coulomb diamond of benzene at different gate
voltages, one corresponding to the center and two towards the charge
degeneracy points with $N=6$ and $N=8$. Also here, the {\it
lineshape} at the inelastic cotunneling threshold has a marked
dependence on the gate-voltage arising from a pronounced gate-voltage asymmetry in the cotunneling amplitudes. Closer to the $N=6$ diamond, virtual tunneling-out processes are closer to resonance (have a smaller energy denominator) and closer to the $N=8$ diamond virtual tunneling-in processes dominate. As for the TD, the monotonously increasing conductance  closer to the $N=8$ diamond is clearly seen to also have a larger step-height.

\section{Conclusions}\label{sec:conclusions}
In conclusion, we investigated cotunneling phenomena in complex
quantum dot systems and demonstrated that systems with weakly broken
degeneracies can exhibit a marked gate-voltage dependence of
the nonlinear cotunneling conductance traces. The effect relies on the
non-equilibrium population of the excited state and is therefore most
pronounced in devices coupled symmetrically to source and drain
electrodes. The inelastic cotunneling threshold was found to be modulated so as to become either cusped or monotonously
increasing, depending on whether the strongest transport channel is via the
ground state or via the first excited state.

In Ref.~\cite{Holm08}, the inelastic cotunneling thresholds were shown to acquire a gate-dependence due to the difference in tunneling-induced level-shifts for the two different levels involved. Whereas that effect shows up with only tunnel coupling to a single lead, it is important to recognize that the gate-dependent modulation of the step which we discuss here relies entirely on the coupling to two different leads. Entering the Kondo regime for which such level-shifts become important, both effects could be observed simultaneously and therefore it would be interesting to study this stronger coupled regime more closely in future studies.

From the technical point of view, we have demonstrated that the widely used simplification of the $T$-matrix approach {(\AppI)} is indeed in {\it quantitative} agreement with the exact fourth order perturbation theory {(GME)} in regions of gate and bias-voltage for which sequential tunneling resonances are strongly suppressed. With a poorer separation of energy scales, i.e. when the inelastic cotunneling threshold is no longer much smaller than the charging energy, \AppI is insufficient but approximation \AppII and the $T$-matrix approach still perform very well and have a fairly large range of validity, within which they yield good agreement with the GME results. In particular, they describe very well the lineshape of the inelastic cotunneling conductance for systems with weakly broken degeneracies, i.e., with $\Delta\ll E_C$. \AppI gives rise to substantial simplifications and allows writing occupation numbers and current in closed analytic form, despite the potential complexity of
the underlying quantum dot systems which is now wrapped up in the virtual transition-amplitudes comprising the effective exchange-cotunneling matrix-elements, see Eq.~(\ref{eq:hint}). As such, \AppI can also be used for the investigation of Kondo-enhanced inelastic cotunneling~\cite{Paaske06} in more complex quantum dot or single-molecule systems, where the most relevant terms in higher order perturbation theory will be the log-singular terms underlying the Kondo effect.

We thank Andrea Donarini for fruitful discussions. We acknowledge
financial support by the DFG within the research programs SPP 1243
and SFB 689.

\end{document}